\documentclass[footinbib,
 reprint,
showpacs,
 amsmath,amssymb,
 aps,
prl,
]{revtex4-2}
\usepackage{mathtools}
\usepackage[all,matrix,arrow,curve]{xy}
\usepackage[utf8]{inputenc}

\def\la{\label}

\usepackage[pdftex]{graphicx}
\usepackage{dcolumn}
\usepackage{bm}

\def\be{\begin{equation}}
\def\ee{\end{equation}}
\def\ba{\begin{eqnarray}}
\def\ea{\end{eqnarray}}

\def\h{h} 
\def\Ham{{\mathrm{Ham}}}
\def\rd{\mathrm{d}}

\def\CG{{\cal G}}
\def\CH{{\cal H}}
\def\CE{{\cal E}}
\def\CF{{\cal F}}
\def\CP{{\cal P}}
\def\CL{{\cal L}}
\def\CQ{{\cal Q}}

\def\CG{{\cal G}}

\def\CM{{\cal M}}

\let\G=\Gamma

\newcommand*{\R}{{\mathbb R}}

\begin{document}
\title{BFV extensions for mechanical systems with Lie-2 symmetry}


\author{Aliaksandr Hancharuk and Thomas Strobl}
\affiliation{Institut Camille Jordan,
Universit\'e Claude Bernard Lyon 1 \\
43 boulevard du 11 novembre 1918, 69622 Villeurbanne cedex,
France
}%
\begin{abstract}
We consider mechanical systems on $T^*M$ with possibly irregular and reducible first class contraints linear in the momenta, which thus correspond to singular foliations on $M$. According to a recent result, the latter ones have a Lie-infinity algebroid $(\CM,Q)$ covering them, where we restrict to the case of Lie-2 algebroids. 
We propose to consider $T^*\CM$ as a potential BFV extended phase space of the constrained system, such that the canonical lift of the nilpotent vector field $Q$ yields automatically a solution to the BFV master equation. We show that in this case, the BFV extension of the Hamiltonian, providing a second corner stone of the BFV formalism, may be obstructed. We identify the corresponding complex governing this second extension problem explicitly (the first extension problem was circumvented by means of the lift of the Lie-2 algebroid structure). We repeatedly come back to the example of angular momenta on $T^*\R^3$: in this procedure, the standard free Hamiltonian does not have a BFV extension---while it does so on $T^*(\R^3 \backslash \{0 \})$, with a relatively involved ghost contribution singular at the origin.


\end{abstract}


\maketitle

\section{Introduction}
One of the main assumptions in textbooks about constrained systems---see, e.g., \cite{HT}---is that the constraint functions $G_a$ are regular on the unconstrained symplectic manifold. Important existence theorems about BRST-BV \cite{BV1,BV2} or BFV  \cite{BFV1,BFV2} extensions have then been proven in this setting only---see, however, \cite{Kazdhan-Felder,MuellerLennart, Herbig} for a notable exception. 
Physically realistic systems often do not satisfy the regularity conditions. 
In realistic situations, group actions can have fixed points and there the corresponding constraints are not regular. And in non-abelian Yang-Mills gauge theories on compact spaces, for example,  the constraints become non-regular at reducible connections. 


On the other hand, without irreducibility and, in particular, regularity assumptions, often the  BV and BFV extensions become hard mathematical problems. Let us illustrate this statement in the finite dimensional setting: Suppose you are given a set $v_a$ of vector fields on a manifold $M$ satisfying 
\be [v_a,v_b] = C^c_{ab} v_c \la{v}
\ee 
for some functions $C^c_{ab}$. Then the corresponding "BRST differential'' $Q$ is readily written down as \be Q = \xi^a v_a - \tfrac{1}{2} C^a_{bc} \, \xi^b \xi^c \tfrac{\partial}{\partial \xi^a} \la{Q}
\ee 
where $\xi^a$ are the "ghosts'',  odd coordinates on an extended graded manifold. At first sight, a  short calculation, using  \eqref{v} together with the Jacobi identity for the Lie bracket of  vector fields,  seems to establish  $Q^2=0$. However, if the vector fields $v_a$ are not linarly independent at each  $x\in M$, there exist functions $t^a_I$ such that 
\be  t^a_I \, v_a \equiv 0 .  \la{t}
\ee 
This happens already for the standard three vector fields $L_a$ generating rotations in $\R^3$ where $x^a L_a  \equiv 0$. Then the structure functions $C^a_{bc}$ entering \eqref{v} are not unique: 
\be C^a_{bc} \sim C^a_{bc} + t^a_I B^I_{bc} . \la{equiv}
\ee 
 The question if, within this equivalence class, there \emph{exists} a choice of the structure functions $C^c_{ab}$ such that $Q$ as defined in \eqref{Q} 
squares to zero is, by the work of Vaintrob \cite{Vaintrob}, equivalent to the question if the singular foliation generated by the vector fields $v_a$ comes from a Lie algebroid covering it.\footnote{In the case of the angular momenta $L_a$, there is a preferred choice of structure functions, namely the constant ones. They can be identified with the structure constants of the Lie algebra $so(3)$. Indeed this example of rotations comes from a simple Lie algebroid, namely the action Lie algebroid $so(3)\times \R^3$.} And this question is, as of this day, an unsolved mathematical problem. 

In the present paper, we want to consider a Hamiltonian constrained system defined on a cotangent bundle $T^*M$, with first class \cite{Dirac} constraints $G_a$  linear in the momenta and a Hamiltonian quadratic in them. Linear constraints are of the form
\be G_a(x,p) \equiv v_a^i(x) p_i \, . \la{Ga}\ee
So we can identify them with the vector fields $v_a$ on $M$ above, since the first class property of the constraints $G_a$ is tantamount to precisely \eqref{v}. We do not restrict to merely regular constraints, as mostly done  in \cite{Ikeda-Strobl1}---otherwise one deals with regular foliations (on $M$ as well as on $T^*M$)---but we permit leaves of different dimensions a priori. 
A quadratic Hamiltonian has the form
\be \mathrm{Ham}= \tfrac{1}{2} g^{ij}(x) p_i p_j  .\la{H}\ee
We assume the coefficient matrix $g^{ij}$ to be non-degenerate,  thus $\Ham$ is the standard Hamiltonian corresponding to a metric $g$ on $M$. We require the usual compatibility of the constraints with the dynamics,  
 \be \{G_a , \mathrm{Ham} \} = \omega_a^b \, G_b  \la{GH}\ee 
for some functions $\omega_a^b$ on $T^*M$. This equips the base manifold $M$ with the structure of a singular Riemannian foliation \cite{KS2,Ikeda-Strobl1,Nahari-Strobl}.

We assume here  that \eqref{t}  captures all dependencies of the vector fields and  there are  no dependencies between the functions $t_I^a$ on $M$. In more mathematical terms, this means that one has a sequence of vector bundles 
\be 
\quad 0 \to F \stackrel{t}{\longrightarrow} E \stackrel{\rho}{\longrightarrow} TM \, , \label{exact}
\ee 
such that $0 \to \Gamma(F) \to \Gamma(E) \to \CF \to 0$ is exact. Here $\CF$ denotes the subset of vector fields obtained as the image of $\rho$, a generating set of which corresponds to the vector fields $v_a=\rho(e_a)$. Likewise, the map $t$ gives rise to the functions $t^a_I$ after the choice of local bases in $e_a$ and $b_I$ in $E$ and $F$, respectively. 

And now the situation is different: Under these assumptions, it has been proven recently \cite{Linf} that there always  exists a homological vector field $Q$, an extension of \eqref{Q}, which describes a Lie 2-algebroid covering the singular foliation generated by the vector fields $v_a$. 

Note that  while  \eqref{t} yields the reducibility  conditions  \be
t_I^a G_a \equiv 0 , \label{dependencies} \ee the above does not imply automatically that there are no further reducibilities between the constraints \eqref{Ga}. This is because to find all  reducibilities, we need to find all independent functions $t_I^a$ satisfying \eqref{dependencies} on $T^*M$---and not just on $M$. 

Consider, e.g., $G_a = \varepsilon_{abc} x_b p_c$, angular momenta. Then besides  $x^a G_a \equiv 0$,  we have the additional dependency $p_a G_a \equiv 0$. But furthermore, there are dependencies between these dependencies and in the end one finds an infinite set of  reducibility conditions leading to an \emph{infinite}  tower of ghosts for ghosts in its  BFV  description. We will illustrate the beginning of this procedure at the  end of this article and provide a complete description elsewhere. 
In  principle, having such a complete description of the tower of dependencies  between  the constraints \eqref{Ga} on $T^*M$, one can construct a Koszul-Tate resolution, which then guarantees the existence of a BFV extension also in the singular context. However, this  description can be very  cumbersome for practical purposes apparently and one may be interested in potential shortcuts. 

As such the following procedure proposes itself: Let us return to the setting described by \eqref{exact}. As mentioned above, there is a homological vector field $Q$ associated  naturally to    \eqref{exact}. If we consider its Hamiltonian lift to the cotangent bundle of the graded space underlying $Q$,  we get a function $\CQ_{BFV}$ of degree one. Due to $Q^2\equiv \tfrac{1}{2}[Q,Q]=0$, this function satisfies automatically the BFV master equation
\be (\CQ_{BFV},\CQ_{BFV})_{\scriptscriptstyle{BFV}} = 0. \la{master}
\ee 
Also it satisfies some other necessary requirements posed on the BFV charge, like the correct appearance of the constraints \eqref{Ga}  and the redundancies \eqref{t} at the lowest  orders in the ghosts. We may thus consider regarding $\CQ_{BFV}$ as the BFV charge of our constrained system. 

A price to pay is that now, in general, it is no more guaranteed that the $\CQ_{BFV}$-cohomology describes  the functions on the (possibly singular) reduced phase  space correctly. There is another famous example, however, where such a deficiency is normally disregarded: The AKSZ formalism \cite{AKSZ1,AKSZ2,AKSZ3} yields a BV extension of  the Poisson sigma model  or  the Chern-Simons gauge theory likewise in a much faster way than when following the usual extension procedure. If  one compares its cohomology at degree zero with gauge invariant functionals modulo the Euler-Lagrange equations, however, the cohomology one finds in the AKSZ framework is in general too large for non-trivial topologies of the source manifold.  To see this explicitly, consider an abelian Chern-Simons theory on a 3-manifold $\Sigma$ where the space of solutions to the Euler-Lagrange equations consist of closed 1-forms and gauge symmetries correspond to adding exact 1-forms to them, thus leading to simply $H^1_{dR}(\Sigma)$, and compare  this  to the results about the corresponding AKSZ-cohomology at degree zero in  \cite{Bona-Kotov}.  

We describe the construction of the BFV phase space obtained in this way  in more detail in the subsequent section. However, there 
 is---in contrast to the Lagrangian  BV formalism---one  more essential  ingredient of  the cohomological approach on the  Hamiltonian level: One needs an extension $\CH_{BFV}$ of the Hamiltonian \eqref{H} such that 
\be (\CQ_{BFV},\CH_{BFV})_{\scriptscriptstyle{BFV}} = 0 \, . \la{master2}
\ee 
Equations \eqref{master} and \eqref{master2} are  the two indispensable
 corner stones of the BFV formalism. Under the given assumptions underlying  the shortcut for  the construction  of $\CQ_{BFV}$, 
the existence of the extension of the Hamiltonian is in general not guaranteed. Its  potential obstructions is the main subject  of the present article. 

%
%
%
%
%
%
%
%
%
%
%
%
%
%
%

\section{BFV phase space and charge}
Much as \eqref{Q} defines a Lie algebroid on $E$ \cite{Vaintrob}, every homological vector field $Q$ of degree 1  on \be 
\CM := E[1] \oplus F[2], \la{CM}\ee which is necessarily of the form 
\begin{eqnarray} Q &=& \xi^a \rho_a^i \tfrac{\partial}{\partial x^i} - \left(\eta^I t_I^a  + \tfrac{1}{2} \xi^b \xi^c C_{bc}^a \right) \tfrac{\partial}{\partial \xi^a} \nonumber \\ &&+ \left(\tfrac{1}{6} \xi^a \xi^b \xi^c \h^I_{abc} -\Gamma^I_{Ja} \eta^J\xi^a\right) \tfrac{\partial}{\partial \eta^I} \, ,  \la{Q2}
\end{eqnarray}
 defines a Lie 2-algebroid structure on the vector bundle $E \oplus F$. Here $E$ and $F$ are the vector bundles appearing in  \eqref{exact} and the numbers in the brackets denote the degrees carried by the local fiber-linear coordinates $\xi^a$ and $\eta^I$ on $E$ and $F$, respectively, in the graded description of the Lie 2-algebroid. 

The coefficient functions entering \eqref{Q2} all have some algebraic and/or  geometrical  meaning: 
One way of viewing them is that the terms 
linear in $(\xi,\eta)$ give rise to a complex \eqref{exact}---which in our case we require, in addition, to be exact on the level of sections---, those quadratic in them are 2-brackets, which are not tensorial but satisfy a Leibniz rule, and the cubic one, which is tensorial, is a 3-bracket, which, if non-zero, reflects the fact that the 2-bracket between sections of $E$ then does not satisfy the Jacobi identity, giving rise  to a Lie 2-algebra. For some introduction to Lie 2-algebras see \cite{Baez2}, for the more general $L_\infty$-algebras see \cite{Lada1,Lada2}. Alternatively, we may view, e.g., the coefficients $\Gamma^I_{Ja}$ as a local expression for an $E$-covariant derivative on $F$. The tensor $\h \in \Gamma(F \otimes \Lambda^3 E^*)$ then satisfies ${}^E\mathrm{D} \h=0$, where ${}^E\mathrm{D}$ denotes the corresponding exterior covariant $E$-derivative. 

There is a vast literature on the geometry of Lie algebroids, see, e.g., \cite{LA1,LA2,LA3}. For further details about the geometry of Lie 2-algebroids see  \cite{Gruetzmann,Goettingen}. For a general  Lie 2-algebroid, the complex  \eqref{exact} is not exact (also not on the level of sections). Under the condition of exactness of  the sequence on the  level sections, thus  providing a resolution  of the  $C^\infty(M)$-module of the vector fields generated by $v_a$, i.e.\ by the image of $\rho \colon  E \to TM$, the existence of a Lie 2-algebroid structure extending  a given sequence \eqref{exact} was  proven in  \cite{Linf}. As mentioned, this is tantamount to the existence of a vector field \eqref{Q2} on the corresponding $\CM$ of \eqref{CM} which squares to zero.

Given these data, we now construct the BFV phase space as the cotangent bundle $T^*\CM$ of \eqref{CM}. This extends the classical variables $(x^i,p_i)\in T^*M$ by a ghost $\xi^a$ (odd and of degree one) for each of the constraints and a ghost-for-ghosts $\eta^I$ (even and of degree two) for each of the dependencies \eqref{dependencies}. Both of the latter two ghost families are accompanied by their momenta of opposite degrees. All this comes together with the BFV symplectic form 
\be \omega_{{\scriptscriptstyle{BFV}}} = \rd x^i \wedge \rd p_i + \rd \xi^a \wedge \rd \pi_a + \rd \eta^I \wedge \rd \CP_I \, . \la{omegaBFV}
\ee 
It is of degree zero and thus so also the graded Poisson bracket $( \cdot , \cdot )_{{\scriptscriptstyle{BFV}}}$ it induces, which takes the following form:
\begin{equation*}
\lbrace p_i, x^j \rbrace = \delta^j_i, \enspace \lbrace \pi_a, \xi^b \rbrace = \delta^b_a,  \enspace \lbrace \CP_I, \eta^J \rbrace = \delta^J_I.
\end{equation*}

Regarding the vector field \eqref{Q2} as a function on the BFV phase space
$T^*\CM$ (by replacing the derivatives simply by the corresponding momenta), one has \begin{eqnarray} \CQ_{BFV} &=& \left(\xi^a \rho_a^i \right) p_i - \left(\eta^I t_I^a  + \tfrac{1}{2} \xi^b \xi^c C_{bc}^a \right) \pi_a \nonumber \\ &&+ \left(\tfrac{1}{6} \xi^a \xi^b \xi^c \h^I_{abc} -\Gamma^I_{Ja} \eta^J\xi^a\right) \CP_I \, ,  \la{QBFV}
\end{eqnarray}
where we use the notation $\rho_a^i \equiv v_a^i$. We notice that, as required by the standard BFV procedure, it indeed extends $\xi^a G_a$ by terms such that \eqref{master} holds true.

The classical constrained system  on $T^*M$ considered here is already uniquely determined by the underlying singular foliation. This is in contrast to the coefficient functions entering \eqref{QBFV}: they are constrained by $Q^2=0$, but, at each step of an extension, they are not unique. This should be also reflected on the BFV level: Indeed, e.g.\ a change \eqref{equiv} of the almost Lie bracket on $E$  can be obtained by lifting the degree preserving diffeomorphism 
\be  \eta^I \mapsto \eta^I  + \tfrac{1}{2}B^I_{ab} \xi^a \xi^b \la{B}\ee
from $\CM$ to $T^*\CM$. Such a change of the structure functions $C^a_{bc}$ does not come for free: also other quantities entering $\CQ_{BFV}$ are then changed correspondingly. E.g., $\h$ receives an additive contribution  by the $E$-covariant exterior derivative of $B \in \Gamma(F \otimes \Lambda^2 E^*)$, $\h \mapsto \h + {}^E\mathrm{D} B + \ldots$, where the dots denote terms quadratic in $B$ (see \cite{Gruetzmann}).


\section{Geometrical interpretation of  $\CH_{BFV}$}
We first observe that the BFV bracket decreases the polynomial degree $\mathrm{pol}$ of  momenta $(p_i,\pi_a,\CP_I)$ by one. Since, in addition, both $\CQ_{BFV}$ and $\Ham$ are homogeneous with respect to $\mathrm{pol}$, we may assume always that the extension $\CH_{BFV} = \Ham + \ldots$ satisfies $\mathrm{pol}(\CH_{BFV})=2$. Identifying momenta with vector fields, we thus may view $\CH_{BFV}$ as a graded-symmetric 2-vector field on \eqref{CM}, 
\be H \equiv H_{BFV} \in \Gamma(S^2 T \CM)  \la{S2}
\ee 
and re-interpret the second master equation \eqref{master2} as 
\be \CL_Q  H = 0 \la{QH}\ee
where $Q=Q_{Lie2} \in \Gamma(T \CM)$. For later use, we remark that a basis of  $\Gamma(S^2 T \CM)$ is spanned by 
\be \tfrac{\partial}{\partial \eta^I} \tfrac{\partial}{\partial \eta^J} , \:   \tfrac{\partial}{\partial \eta^I}\tfrac{\partial}{\partial \xi^a}, \:  \tfrac{\partial}{\partial \eta^I} \tfrac{\partial}{\partial x^i}, \:  \tfrac{\partial}{\partial \xi^a} \tfrac{\partial}{\partial \xi^b}, \:  \tfrac{\partial}{\partial \xi^a} \tfrac{\partial}{\partial x^i}, \:  \tfrac{\partial}{\partial x^i} \tfrac{\partial}{\partial x^j} \la{basis}
\ee  
where a graded symmetrization is understood and the degrees are $-4$, $-3$, $-2$, $-2$, $-1$, and $0$, respectively. $H$ being of degree zero, one needs coefficient functions on $\CM$ of the respective opposite degrees when using this basis. A graded manifold equipped with a degree 1 vector field is called a Q-manifold \cite{Schwarz}; if there is, in addition, \eqref{S2} of degree 0  satisfying \eqref{QH}, we call it an $HQ$-manifold. 

Already an $H$-structure alone, that is a degree zero tensor \eqref{S2} without any compatibility conditions, contains interesting geometrical data, at least if the coefficients of the very  last term  in \eqref{basis} form a non-degenerate matrix---in which case we then call the $H$-structure non-degenerate. It then defines a metric $g$ on $M$ (subject to no further conditions). But this is not all: Consider the terms linear in $p_i\sim \partial_i$. Due to the non-degeneracy of $g^{ij}$, we may absorb  all such terms by completing the square with a redefinition of the momenta $p_i \to p_i^\nabla$, where 
\be p_i^\nabla := p_i - \Gamma^a_{bi} \xi^b \pi_a - (\Gamma^I_{Ji} \eta^J   - \tfrac{1}{2}\gamma_{abi}^I\xi^a\xi^b) \CP_I\, . \la{pgamma}
\ee 
Then the 2-tensor \eqref{S2}, rewritten as a function $\CH$ in  $C^\infty(T^*\CM)$, takes the form
\be \CH = \tfrac{1}{2} g^{ij} p_i^\nabla p_j^\nabla  + \tfrac{1}{2} \Sigma_{(2)}^{ab}  \pi_a \pi_b  + \Sigma_{(3)}^{aI}  \pi_a \CP_I  
 + \tfrac{1}{2}  \Sigma_{(4)}^{IJ}   \CP_I \CP_J \la{expansion}
\ee 
where $\Sigma_{(k)}^{\cdot \cdot}$ are functions on $\CM$ of degree $k$. 
Thus, for example,
\be \Sigma_{(2)}^{ab} = \tfrac{1}{2}\Sigma_{cd}^{ab}\xi^c\xi^d + \Sigma_{I}^{ab} \eta^I \label{Sigmadecomp}\ee 
for appropriate tensors $\Sigma_{cd}^{ab}$ and  $\Sigma_{I}^{ab}$. 
The coefficients in \eqref{pgamma} have a geometrical interpretation: Consider a coordinate transformation $\xi^a \mapsto M^a_b(x) \xi^b$ (all other coordinates unchanged). Lifting it to $T^*\CM$  so as to leave \eqref{omegaBFV} invariant---or, equivalently, considering its tangent map---yields $\pi_a \mapsto (M^{-1})^b_a \pi_b$ together with $p_i \mapsto p_i + (M^{-1})^a_c M^c_{b,i} \xi^b \pi_a$ (all other momenta unchanged). Requiring \eqref{pgamma} to 
remain form-invariant under such transformations shows that $\Gamma^a_{b} = \Gamma^a_{bi} \rd x^i$ are the local 1-forms representing a connection $\nabla$ on $E$. Similarly, $\Gamma^I_{J}= \Gamma^I_{Ji} \rd x^i$ corresponds to a covariant derivative $\nabla$ on $F$. $\gamma \in \G(F \otimes \Lambda^2 E^* \otimes T^*M)$ is needed, finally, to also establish form covariance of \eqref{pgamma} w.r.t.\ \eqref{B}; it is part of a connection in a graded sense. 

An almost Q-structure (a degree one vector field $Q$ which does not necessarily square to zero) equips the graded manifold \eqref{CM} with the structure of a (not necessarily exact) sequence of vector bundles as in \eqref{exact} together with 2-brackets and one 3-bracket. A non-degenerate H-structure equips it in the lowest order with a metric $g$ on the base and in the next order---the terms linear in the momenta $p_i$ (see also   \eqref{pgamma})---with a connection defined on all of \eqref{exact} (note that we can always equip $TM$ with the canonical Levi-Civita connection  of $g$, moreover).

Given an almost HQ-structure, where we have both,  an almost  Q-structure $Q$ and  an $H$ as above, \eqref{QH} determines compatibility conditions.
 To obtain tensorial formulas for these  on the nose, it is useful to re-express \eqref{QBFV} in terms of the variables \eqref{pgamma}. Then, e.g., the new coefficient of the term quadratic in $\xi$ is $-C^a_{bc}+ \rho^i_b \G^a_{ci} - \rho^i_c \G^a_{bi}=: {}^ET^a_{bc}$ and has the geometric interpretation of what is called an $E$-torsion ${}^ET\in \G(E \otimes \Lambda^2E^*)$ \cite{KS2}, generalizing the ordinary torsion on a tangent bundle to $E$. The bracket of two covariant vector fields \eqref{pgamma} yields the curvatures of $\nabla$, both on $E$ and $F$ (as well as a contribution proportional to $\mathrm{D}\gamma$). To lowest orders we find:
\ba \CL_Q H &=& \tfrac{1}{2} \xi^a \left({}^E \nabla_{\! e_a} \, g\right)^{ij} \partial_i^\nabla \partial_j^\nabla \nonumber \\ && -\tfrac{1}{2} \xi^b \xi^c \left( S^a_{bci} - t^a_I \gamma^I_{bci} - \Sigma^{da}_{bc}\rho^k_d g_{ki}\right) g^{ij} \partial_j^\nabla \tfrac{\partial}{\partial \xi^a} \nonumber \\ &&   + b^I \left( t^a_{I;j}g^{ji}  + \Sigma^{ba}_I \rho_b^i \right)\partial_i^\nabla \tfrac{\partial}{\partial \xi^a} + \ldots  \, .
\la{firstterms}
\ea  
Here $\partial_i^\nabla$ is the vector field corresponding to \eqref{pgamma} and ${}^E \nabla$ denotes the $E$-covariant derivative acting on $TM$ according to ${}^E \nabla_{s} v = [\rho(s), v] + \rho(\nabla_v s)$, where $s \in \G(E)$ and $v \in \G(TM)$. $\Sigma_{(2)}^{ab}$ has been decomposed according to \eqref{Sigmadecomp},
a semicolon denotes a covariant derivative,
and we introduced the abbreviation
\be S^a_{bci} =  {}^ET^a_{bc;i} + \rho_b^j R^a_{cji} - \rho_c^j R^a_{bji}
%
. \la{S}
\ee 
The tensor $S\in \Gamma(E \otimes \Lambda^2 E^* \otimes T^*M)$ measures the compatibility of the 2-bracket and the connection $\nabla$ on $E$, see \cite{KS2}. As for Jacobi identities of the brackets in a higher Lie algebroid, also here we do not expect \eqref{QH} to ensure $S$ to vanish on the nose, but to instead do so up to some appropriate boundary term only; what this means precisely is subject of the next section. 

Let us mention that the current method is efficient in obtaining Bianchi type of identites for quantities such as $S$.  $Q^2=0$, which we have when $Q$ comes from a Lie 2-algebroid, implies that the application of another $\CL_Q$ to  \eqref{firstterms} vanishes identically. From this one can read off, for example, that \eqref{S} satisfies 
\be {}^E \mathrm{D} S - \nabla \langle t, \h \rangle_{\!\scriptscriptstyle{F}}= 0 \, .\label{Bianchi}
\ee

Here ${}^E \mathrm{D}$ denotes the $E$-exterior covariant derivative associated naturally to ${}^E\nabla$: on $TM$ it acts as specified above, on $E$ according to ${}^E\nabla_s s' =[s,s']_E + \nabla_{\rho(s')} s$ for every  $s,s' \in \G(E)$, and on elements in $\Lambda^\bullet E^*$ by a straightforward generalization of the Cartan formula for the de Rahm differential (\cite{Gruetzmann,KS2}). The identity \eqref{Bianchi}---specialized to a Lie algebroid, where $h=0$---was of essential importance in the construction of the BV-extension 
\cite{HPSM} and the above derivation constitutes a significant simplification of the one provided there.

%

\section{Complex governing the extension}
\vspace{-3mm}
In the construction of \eqref{QBFV}, the sequence of vector bundles \eqref{exact} and  its exactness on the level of sections plays a crucial role. There is a similar sequence which governs the extension problem \eqref{H},\eqref{GH},\eqref{master2}, but, despite being constructed 
out of the previous one, this one is in general no longer exact on the level of sections. 

 Let us denote the complex \eqref{exact} by $\CE^\bullet$ (so, in particular, $\CE^0 =TM$, $\CE^{-1} =E$, and $\CE^{-2} =F$) and  tensor it with itself shifted to the right, $ \CF^\bullet := \CE^\bullet\otimes \CE^\bullet[-1]$, where, by definition, $\CE^i[-1] =\CE^{i-1}$.  At degree zero, e.g., one has $\CF^0 = TM \otimes E \oplus E \otimes TM$, where in the  first term both factors carry degree 0, while in the second one, elements in $E$ enter with degree -1 and those in $TM$ with degree +1, again adding up to 0.

By a standard construction (see, e.g., \cite{product}), $\CF^\bullet$ is again a complex. 
It is, however,  not yet the sequence $\CG^\bullet$ we are interested in:
\begin{widetext}
\be \CG^\bullet := 0 \stackrel{\delta}{\longrightarrow} S^2 F \stackrel{\delta}{\longrightarrow} F \otimes E \stackrel{\delta}{\longrightarrow} F \otimes TM \oplus \Lambda^2 E \stackrel{\delta}{\longrightarrow}E \otimes TM \stackrel{\delta}{\longrightarrow} S^2 TM\, , \la{CG}\ee
\end{widetext}
where the degrees are  such that $\CG^0 = S^2 TM$, $\CG^{-1} = E\otimes  TM$, etc. Now one observes that $ \CG^\bullet \xhookrightarrow{} \CF^\bullet[1]$ (essentially it is embedded as a graded symmetrization).
 For example, typical elements $\Phi \in \CF^0$ and $\varphi \in \CG^{-1}$ are of the form $\Phi = \Phi^{ia} \, \partial_i \otimes e_a + \bar{\Phi}^{ai} \, e_a \otimes \partial_i$  and $\varphi =\varphi^{ai} \, e_a \otimes \partial_i$, respectively, and then $\CG^{-1}$ is embedded into $\CF^0$ by the diagonal map, $\Phi^{ia}:= \varphi^{ai}$, $ \bar{\Phi}^{ai}:=\varphi^{ai}$. The codifferential $\delta$ is easily identified with 
  \be \delta :=  \rho_a^i p_i \tfrac{\partial}{\partial \pi_a} + t_I^a \pi_a \tfrac{\partial}{\partial \CP_I} \la{delta}
\ee 
when replacing the vector fields \eqref{basis}, which provide a basis for   $\CG^\bullet$,  by the corresponding momenta. 

In general, for the cohomology of a tensor product of two complexes there is a K\"unneth formula. It says that even if the cohomology of each of the complexes is trivial, which is the case here, there can still be a non-zero contribution called ``torsion''. Below we will  provide an example that, in general, $\CG^\bullet$ is not exact, even not on the level of sections. 

It is remarkable that \eqref{delta} also generates the coboundary operator of the complex \eqref{exact}. In fact, if all the independent  relations between the constraints $G_a = \rho^i_a p_i$ on $T^*M$ are provided by the functions  $t^a_I(x)$  (these are all such dependencies that depend on $x$ only, but there could be further ones that also depend on momenta $p$ in principle), then the operator \eqref{delta} acting on $C^\infty(T^*\CM)$ would be  the Koszul-Tate differential  \cite{HT}. And in that case, it would have no cohomology on $C^\infty(T^*\CM)$  and neither  so when  acting on \eqref{CG}, which is the restriction of $(C^\infty(T^*\CM),\delta)$ to functions quadratic in the momenta---such as the complex \eqref{exact} can be identified with its restriction to the subcomplex linear in the momenta. 
But in general, the true Koszul-Tate complex is much bigger, as we will illustrate by means of an example at the end of this article. 

The main point here is, however, that it is precisely the cohomology of \eqref{CG} that governs the extendability problem $\CH_{BFV} = \Ham + \ldots$ satisfying \eqref{master2} for $\CQ_{BFV}$ as in \eqref{QBFV}.

\emph{If}  $\G(\CG^\bullet)$ is exact, the existence of the BFV extension $H_{BFV}$ is guaranteed. This follows from a standard consideration:   $(\CQ_{BFV}, \cdot )  = \delta + X_{rest}$. At each step when adding terms from the right to the left in \eqref{CG}, one finds an expression that is already $\delta$-closed. Now, $\delta$ having no cohomology, one can always add a $\delta$-exact contribution from the next level so as to cancel it. In general, the resulting expression for $H_{BFV}$ will contain all possible terms compatible with degrees, as is also the case for \eqref{QBFV}; together they then define an HQ-structure on \eqref{CM}. 

The converse is certainly not true: The complex $\G(\CG^\bullet)$ can  have 
non-trivial cohomology, but the extension problem for a particular Hamiltonian \eqref{H} may still lead to  trivial cohomology classes and thus be unobstructed.

The absence of a cohomology also leads to remarkable, purely geometrical formulas. To illustrate this, assume that 
\be  H^{-1}(\G(\CG^\bullet),\delta)=0. \la{H1}
\ee 
Then for every $\varphi \in \Gamma(\CG^{-1})$ such that $\delta \varphi =0$,  there is some $\psi \equiv \Psi^{Ii} b_I \otimes \partial_i + \tfrac{1}{2}M^{ab} e_a \wedge e_b \in \Gamma(\CG^{-2})$ such that
 $\varphi = \delta \psi$. Concretely,
\be  \rho_a^i \, \varphi^{aj} +  \rho_a^j \,  \varphi^{ai} = 0 \quad \Rightarrow \quad  \varphi^{ai} = t^a_I \Psi^{Ii} + \rho_b^i M^{ab} \, ,\la{useful}
\ee
where $M^{ab}(x) = -M^{ba}(x)$. The exactness at a given degree is preserved if one tensors a complex with a fixed $C^\infty(M)$-module; 
this permits adding spectator indices to the quantities in \eqref{useful}. 

As a first application, let us return to the initial compatibility condition \eqref{GH}. It can be shown \cite{Nahari-Strobl}
that \eqref{GH} holds true in arbitrary local patches iff there exists a connection $\nabla$ on $E$ such that its induced $E$-connection annihilates the metric, ${}^E\nabla g = 0$. If $\Gamma^b_{ai}$ denote the connection coefficients of $\nabla$, we can choose $\omega_a^{b}:=\Gamma_{ai}^{b}(x)g^{ij}p_j$. The difference between two connections is a tensor field and then \eqref{GH} implies that this tensor is $\delta$-closed. This provides the following equivalence for the choice of the connection coefficients\be \omega^{ai}_b \sim \omega^{ai}_b +  t^a_I \Psi^{Ii}_b + \rho_c^i M^{ac}_b , \la{equiv2}
\ee 
where $M^{ac}_b=-M^{ca}_b$, corresponding to $\delta$-exact contributions. 
If \eqref{H1} holds true, these are \emph{all} the ambiguities in the choice of the connection on $E$ such that ${}^E\nabla g = 0$.

Even more remarkable is the following decomposition that one finds for the tensor $S$  defined in \eqref{S}. Using a connection $\nabla$ such that ${}^E\nabla g=0$, 
 the geometrical quantity $S$  
is always $\delta$-closed and therefore, assuming \eqref{H1}, it can  be decomposed into some $\gamma \in \Gamma(F  \otimes \Lambda^2 E^*\otimes T^*M)$ and $\Sigma\in \Gamma(\Lambda^2 E \otimes \Lambda^2 E^*)$ as follows:
\be S^a_{bci} = t^a_I \gamma^I_{bci} + \rho_d^j g_{ij} \Sigma^{da}_{bc} \, . 
\la{Sdecomp} 
\ee 
There are, in general, no explicit expressions for these tensors. In fact, they are even not uniquely determined: We can change $\gamma$ and $\Sigma$ simultaneously by
\ba  \Sigma^{ab}_{cd} &\mapsto& \Sigma^{ab}_{cd} + t^a_I \, V^{Ib}_{cd} - t^b_I \, V^{Ia}_{cd} \nonumber \\ 
\gamma^I_{abi} &\mapsto&\gamma^I_{abi} + g_{ij}\rho_c^j  \, V^{Ic}_{ab} \la{deltaV}
\ea 
for any $V \in \Gamma(F\otimes E \otimes \Lambda^2 E^*)$ without changing $S$---and these are all such ambiguities, if also $H_\delta^{-2}(\G(\CG))=0$.


Finally, the identity  $\{t^a_I G_a, H\}$ yields 
\be 
t^a_{I;i} g^{ij} \!\rho_a^k \, p_jp_k =0 \, .  \label{1}
\ee 
Thus, $t^{a;i}_I  \equiv t^a_{I;j} g^{ij}$ satisfies the condition of \eqref{useful} and, if \eqref{H1}  holds true, there exist tensors  $\tilde \Psi$ and $\tilde M$ such that 
\be  t^{a;i}_I = t^a_J \tilde\Psi^{Ji}_I + \rho^i_b \tilde M_I^{ab} \, . \la{2}
\ee 

These considerations can be considered as holding true on the purely geometrical level, without any application to physical models. 

On the other hand, as explained above, the $\delta$-cohomology governs the physical extension problem here. For example, the second line in \eqref{firstterms} says that $S$ needs to be $\delta$-exact (and, by a general feature of the procedure, also shows implicitly that it is already  $\delta$-closed), $S=\delta(\gamma + \Sigma)$, where $\gamma$ and $\Sigma$ are the quantities entering $\CH$, see \eqref{pgamma}, \eqref{expansion}, and \eqref{Sigmadecomp}. In deriving \eqref{Sdecomp} we only demanded $S$ to be $\delta$-exact with respect to some tensors, but in a slight abuse of notation we denoted them already by $\gamma$ and $\Sigma$, the quantities we need to choose for the expansion  \eqref{expansion}. We have not done so in \eqref{2}, which can be rewritten as $\nabla t = \delta(\tilde\Psi+\tilde M)$ and one may wonder where the analogue of the first term on its right hand side is in comparison to the last line in \eqref{firstterms}, while evidently $\tilde M_I^{ab}$ can be identified with $\Sigma_I^{ab}$. In fact, this depends on  the connection on $F$: if one uses a different connection than the one appearing in \eqref{pgamma}, then their difference is a tensor,  which can be identified with  $\tilde\Psi^{Ji}_I$. So also the last line in \eqref{firstterms} says that $\nabla t$ should be $\delta$-exact.

\section{Examples  with $t=0$}
The prototype of a singular constraint $T^*\R \cong \R^2$ is $G=x p$. It is singular at the origin, the constraint surface has the shape of a cross. The corresponding singular foliation on $M=\R$ is generated by  $v = x \partial_x$ or, by putting $E=\R \times M$ with an anchor such  that $\rho(1)=v$. The bundle $F$ then has rank zero and the exact sequence is of length one only. The corresponding BFV charge is simply $\CQ_{BFV}= \xi x p$.

There is no compatible non-degenerate Hamiltonian coming from a metric $g$ in this case, since $\R$ with a non-regular leaf structure does not admit a  singular Riemannian foliation. 
However, if we start with a Hamiltonian $\Ham = \tfrac{1}{2} h(x) p^2$, dropping the condition that $h = 1/g$, then a BFV extension exists as long as $h$ vanishes at least quadratically near the origin; it then takes the form: $\CH_{BFV} =  \tfrac{1}{2} h(x) p^2 + x^2 \left( \tfrac{h}{x^2}\right)' \xi \pi p$.  

More generally, if $t=0$, the anchor map $\rho \colon E \to TM$ needs to be injective on a dense open subset of $M$: this follows from injectivity of the induced map from $\G(E)$ to  $\G(TM)$  when $t=0$ (see the exactness condition expressed in the sentence containing \eqref{exact}). And if one insists on a non-degenerate metric $g$ satisfying \eqref{GH}, then $t=0$ implies injectivity of the anchor map $\rho \colon E \to TM$ everywhere. 
This in turn implies that the foliation on $M$ and the constraints \eqref{Ga} on $T^*M$ are regular.

Regular foliations can still yield geometrically interesting examples. To illustrate this, consider even $E=TM$, $\rho = \mathrm{id}$. While the BFV charge is very simple in this case, $\CQ_{BFV} = \xi^i p_i$, non-trivial  geometry remains in the BFV extension of the Hamiltonian: To satisfy the compatibility \eqref{GH}, we need a connection $\nabla$ on $TM$ such that \be g_{ij;k} +T_{ijk} +T_{jik}=0, \la{metricity}\ee 
where $T_{ijk}=g_{il}T^l_{jk}$; here  $T^l_{jk}$ denote the components of the torsion tensor. The Levi-Civita connection satisfies this condition evidently. The ambiguity \eqref{equiv2} translates into the freedom of the choice of the torsion of the connection, but maintaining \eqref{metricity} instead of metricity. One then finds
\be \CH_{BFV}=\tfrac{1}{2}g^{ij} p_i^\nabla p_j^\nabla + \tfrac{1}{2} R^k{}_{ij}{}^l\xi^i\xi^j \pi_k\pi_l + \tfrac{1}{4} T^k{}_{ij}{}^{;l}\xi^i\xi^j \pi_k\pi_l  \la{T}
\ee 
where $R^i{}_{jkl}$ denote the components of the Riemann tensor. 
If we choose the Levi-Civita connection for $\G^i{}_{jk}$ in $p_i^\nabla = p_i + \G^j{}_{ki}\xi^k\pi_j$, then \eqref{metricity} is satisfied identically and the last term in \eqref{T} disappears, but one is still left with the curvature term.

Inspection of  \eqref{S}  shows that the last two terms in \eqref{T}  combine into the  tensor $S$ for $E=TM$. In \cite{Ikeda-Strobl1}, such a BFV extension was provided under  the assumption that $S$ vanishes. The  extension \eqref{T} of the Hamiltonian generalizes this result, with the interesting contributions of curvature and torsion which combe into the geometrical quantity $S$, which, for $E=TM$, in turn can be identified also with the curvature of  a dual connection, see \cite{KS2}.



\section{Angular Momentum I: \\Example of an obstruction}
Let us now consider the phase space $T^*\R^3$ with constrained angular momentum,  
$G_a = \varepsilon_{abc} x^b p^c$ or $\vec{G}= \vec{x} \times \vec{p}$. Then one has
\be \CQ_{BFV}= \vec{\xi} \cdot (\vec{x} \times \vec{p}) + \tfrac{1}{2} 
 (\vec{\xi} \times \vec{\xi}) \cdot\vec{\pi} - \eta \, \vec{x} \cdot \vec{\pi} \, .\la{QBFVrotations}
\ee
The second term is the (cotangent lift of the) Chevalley-Eilenberg differential of the Lie algebra $so(3)$, the first two terms the Chevalley-Eilenberg differential for $so(3)$ acting on $\R^3$, which  then corresponds to the BRST charge \eqref{Q} of the action Lie algebroid $E=so(3) \times \R^3$. 
The last term takes care of the dependency $\vec{x} \cdot \vec{G} \equiv 0$ of the constraints. It is easy to verify that the expression \eqref{QBFVrotations} satisfies the master equation \eqref{master}.



In this example, $M=\R^3$, $E = \R^3 \times M$, $F =  \R \times M$, and the maps $\rho$ and $t$ in \eqref{exact} can be identified with the sections  $\rho = \varepsilon_{abc} x^b e^a \otimes \tfrac{\partial}{\partial x^c}$ and 
\be t= x^a e_a  \otimes b^* \, , \label{trot}\ee
respectively; here $(e^a)_{a=1}^3$ denotes a basis in $E^*$  and $b^*$ a basis in $F^*$. The kernel of $\rho$ is one-dimensional outside the origin, while it is all of $ E_{\vec{x}=0} =\R^3$ at the origin $0 \in M$. The map $t$ spans the radial vectors in  $\ker \rho$ for all $\vec{x} \neq 0$, but, for continuity reasons, vanishes at $0$. Thus, the complex \eqref{exact} has no cohomology outside the origin, but $H_{\vec{x}=0}^{-1}(\CE,\delta)=\R^3$. And still---again for continuity reasons (every radial vector field has to vanish at the origin)---it is exact on the level of sections: $H^\bullet(\G(\CE^\bullet),\delta)=0$. Correspondingly, the extension problem for the BFV charge \eqref{QBFVrotations} when taking into account (only) the $x$-dependent dependencies of the constraints, has  not been obstructed.

Let us now turn to the extension problem of the Hamiltonian \eqref{H},  the  main subject of this paper, for the standard metric on $M=\R^3$,
\be \Ham = \tfrac{1}{2} \vec{p}\cdot \vec{p} \, . \la{3}\ee  
For this purpose we first equip the bundles $E$ and $F$ with their canonical flat connections (for what concerns $E$, this is also motivated by the fact that \eqref{GH} is satisfied with $\omega_a^b=0$). Then one has
\be \nabla t =  \rd x^a  \otimes e_a \otimes b^* \stackrel{\sim}{=}
\tfrac{\partial}{\partial x^a}  \otimes e_a\otimes b^* \, ,
\ee
where  in the second equality we  used the standard metric of $M$ for the identification, then yielding an element in $\G(E\otimes TM) \otimes_{C^\infty(M)} \G(F^*)$. It is easy to see that $\delta (\nabla t) = 0$ which is equivalent to \eqref{1}: applying $\rho$ to $e_a$ and symmetrizing over the two ensuing entries in $TM$ gives zero due to the contraction with the $\varepsilon$-tensor. 

Now the main observation of this short section: $\nabla t$ cannot be $\delta$ exact, i.e.\ it cannot be of the form \eqref{2}, since both $t$ and $\rho$ vanish at the origin. This shows, on the one hand, that here 
\be H^{-1}(\G(\CG^\bullet),\delta) \neq 0 ,
\ee 
and, on the other hand, that the BFV extension of \eqref{3}  
 is indeed obstructed: As we learn from the last line in \eqref{firstterms}, we \emph{need} $\nabla t$ to be exact for the BFV-extension of the Hamiltonian within the present framework.

Note also that any other choice of connections on $E$ and $F$ would not help, since their contribution to $\nabla t$ vanish at $x=0$ as well. In fact, the cohomology class of $\nabla t$ does not depend on such choices.

\section{Examples with Lie-2 gauge symmetry}
As a technically more involved example, we  truncate the expansion  for the BFV extension of the Hamiltonian (while keeping the general form for $\CQ_{BFV}$ as in \eqref{QBFV}): Putting $\Sigma_{3}$, $\Sigma_{4}$, and $\Sigma^{ab}_I$ to zero in \eqref{expansion}, one obtains\footnote{This example is an improvement of a result of \cite{Ikeda-Strobl1} obtained in the context of Cartan-Lie algebroids, where the $\Sigma$-term in \eqref{beauty} is absent. During the preparation of the present work, we were informed by N.\ Ikeda that he found a similar extension in the context of Courant algebroids, see \cite{Ikedaneu}.} 
 \be \CH_{BFV} = \tfrac{1}{2} g^{ij} p_i^\nabla p_j^\nabla + \tfrac{1}{4} \Sigma_{cd}^{ab} \xi^c\xi^d \pi^\nabla_a \pi^\nabla_b \, , \la{beauty}
\ee 
where $p_i^\nabla$ is given by  \eqref{pgamma} and $\pi_a^\nabla = 
\pi_a - \lambda_{ab}^I \xi^b \CP_I$ for some $\lambda \in \G(\Lambda^2 E^* \otimes F)$. This extension of $\pi_a$ to $\pi_a^\nabla$
 has the effect of 
 making \eqref{beauty} covariant with respect to both, changes of frames as well as \eqref{B}, where then $\lambda \mapsto \lambda + B$. 
Now we learn from  \eqref{firstterms} that, under these assumptions, \emph{necessary} conditions for the validity of \eqref{master2} are
\be  {}^E\nabla g = 0 \: , \quad  \nabla t = 0 \: , \quad \, S = \delta(\gamma + \Sigma). \la{nablat} \ee  
Note that the second condition implies that $t$ needs to have constant rank.
On the other hand, the last condition in \eqref{nablat} is not overly restrictive, since always $\delta S=0$. 

A priori, there are many more conditions to be satisfied for \eqref{master2} to hold, but with the following trick, one may show that they can be all reduced to one only: For this purpose, we first remark that within \eqref{nablat} we can always change the two quantities $\gamma$ and $\Sigma$---which are then to enter the extension \eqref{beauty}---according to \eqref{deltaV}. Since $t$ has a constant rank, we can choose some $C\subset E$ 
such that $E= C \oplus t(F)$. One may now see that the transformations \eqref{deltaV} can be used to assure $\Sigma \in \G(\Lambda^2 C \otimes \Lambda^2 E^*)$. Then the only remaining condition is:
\be {}^E\mathrm{D}_\lambda \Sigma = 0 \, .\la{DSigma}
\ee
Here ${}^E\mathrm{D}_\lambda$ is defined as ${}^E\mathrm{D}$ but replacing the 2-bracket on $E$ by  
$[s,s']_\lambda = [s,s'] + t(\lambda(s,s'))$, where   $\lambda$ is the tensor entering $\pi^\nabla_a$. 

We finally remark that from \eqref{Bianchi} one can conclude an equation similar to \eqref{DSigma}: One first observes that the operators ${}^E\mathrm{D}$ and \eqref{delta} commute, $[{}^E\mathrm{D},\delta]=0$. Using \eqref{Bianchi}, one then finds that $\delta({}^E\mathrm{D}(\gamma+\Sigma)) =  \langle t, \nabla h\rangle_F$, where the $h$-contribution only adds to ${}^E\mathrm{D}\gamma$. If $H^{-2}(\G(\CG^\bullet),\delta)=0$ holds, moreover, one finds that ${}^E\mathrm{D}\Sigma$ is part of a coboundary for some $\tilde V \in \G(\Lambda^3 E^* \otimes E \otimes F)$. \eqref{DSigma} then translates into the condition that  $\tilde V$ needs to be a contraction of $\lambda$ with $\Sigma$.

\section{Angular Momentum II: \\Outside of the spatial origin}
We want to illustrate the above formulas by means of the example of the angular momentum we discussed already before, but this time excluding the origin $\vec{x}=0$. 

On the parts of $T^*\R^3$ where $\vec{x} \neq 0$, the dependency $\vec{x} \cdot \vec{G} \equiv 0$ is in fact already sufficient to describe all the dependencies of the constraints: this is the case since, assuming  $\vec{x} \neq 0$, the constraint surface $ \vec{G} = \vec{x} \times \vec{p} \approx 0$ implies $\vec{p} \approx \lambda \vec{x}$ for some  $\lambda \in \R$. Thus the dependency $\vec{p} \cdot \vec{G} \equiv 0$ is automatically implemented when $\vec{x} \cdot \vec{G} \equiv 0$ is taken care of.

Therefore now, even if one follows the Koszul-Tate procedure, one does not need to introduce an additional ghost of degree two to take into account all the dependencies of the constraints: the BFV charge 
\eqref{QBFVrotations} is the one that gives the correct cohomology of observables when the $\vec{x}$-origin is excluded. This is, on the other hand, not the case, if the last term in \eqref{QBFVrotations} is supressed---as one finds it sometimes in the  literature in the  treatment of mechanical models with rotational invariance in the BRST-B(F)V formalism.

Excluding the $\vec{x}$-origin, there are now no more obstructions to BFV-extend the classical Hamiltonian \eqref{3} to an $\CH_{BFV}$ satisfying the second master equation \eqref{master2}. And, as it turns out, it can even be put into the special form \eqref{beauty}. As we learn from \eqref{nablat}, we need for this that the tensor \eqref{trot} is covariantly constant. Let us choose for this purpose the connections  on $E$ and $F$ by means of 
\be \nabla e_a = -\frac{x_a}{r^2} \, \rd x^b \otimes e_b \label{nablarot}\ee  and $\nabla b = 0$, respectively. Note that the first choice implies that the radial section $x^a e_a$ is covariantly constant, $\nabla (x^a e_a)=0$. Together these choices  indeed yield $\nabla t=0$. It was this condition that we could not satisfy previously; here it is possible also only due to the singularity of the connection $\nabla$ on $E$ when approaching the origin.

One also verifies easily, that this choice of the connection on $E$ guarantees 
${}^E\nabla g = 0$, a necessary condition for the BFV extension even without the restricting $\CH_{BFV}$ to be of the form \eqref{beauty}, see the first line in \eqref{firstterms}. The tensor $S$ looks as follows with the above choices for the connection: \be S = \frac{\epsilon_{bce}x^e}{r^2} e_a \otimes e^b \otimes e^c \otimes   \rd x^a \, .\label{Srot}\ee

$S$ now has a contribution coming from the tensor $\gamma$ in its decomposition \eqref{Sdecomp}, which enters the covariantized momentum \eqref{pgamma}:
\be \vec p^{\nabla} = \vec p - \frac{1}{r^2}\vec \pi (\vec \xi \cdot \vec x) + \frac{\vec x}{2r^4} \vec \xi\cdot(\vec \xi \times \vec x)\CP \, .\ee 
The complete BFV extension of the Hamiltonian then takes the form
\begin{equation}
	\label{Hrot1}
		 \CH_{BFV} = \tfrac{1}{2} \, \vec p^{\nabla} \! \! \cdot \! \vec p^{\nabla} - \frac{1}{4r^4} \bigl[ \vec \xi \cdot (\vec \xi \times \vec x) \bigr] \, \bigl[
		 \vec \pi \cdot (\vec \pi \times \vec x)\bigr] . 
\end{equation}
This corresponds to the following tensor $\Sigma$ in \eqref{beauty}, 
\be \Sigma^{ab}_{cd} = - \frac{\varepsilon_{abe}\varepsilon_{cdg}x^ex^g }{r^4}\, , \ee
which  satisfies the consistency condition \eqref{DSigma} for $\lambda =0$. Together with the above choice for $\gamma$, this completes the decomposition \eqref{Sdecomp}  of \eqref{Srot}, which then satisfies  the last condition in \eqref{nablat}. According to the discussion following \eqref{beauty}, this then proves the validity of \eqref{master2}, which certainly one can also verify directly to hold true for \eqref{QBFVrotations} and  \eqref{Hrot1}.

The extension of \eqref{3} to $\CH_{BFV}$ is not unique, also not for a fixed choice of the BFV charge  \eqref{QBFVrotations}. One can, at each stage, change contributions to $\CH = \CH_{BFV}$ in \eqref{expansion} by $\delta$-exact terms (once more underlining the importance of the complex $\CG^\bullet$ for the given extension problem). For a given Hamiltonian, the first such an ambiguity is provided by the choice of the connection on $E$, which we found in \eqref{equiv2}. It is now not difficult to see that with the choice  $\Psi^{i}_b = \tfrac{1}{r^2}\delta^i_b$, $M^{ac}_b = \tfrac{1}{r^2}\epsilon_{acb}$, the connection coefficients corresponding to \eqref{nablarot} can be made to vanish: $\nabla e_a = 0$ for this modified, equivalent connection. Now the connection on $F$, if taken as the corresponding coefficient in \eqref{pgamma}, is fixed by requiring the vanishing of the last line in \eqref{firstterms}, i.e.\ by requiring $\nabla t = \rho^i_b \Sigma^{ab} e_a\otimes \rd x^i$: One finds 
\be \nabla b =  \frac{x_a \rd x^a }{r^2} \otimes b \, , \ee 
which, in particular, has the property that the section $\frac{1}{r}b$ is covariantly constant,  $\nabla(\tfrac{1}{r}b) = 0$. After the dust clears, one then finds the following, relatively short BFV-Hamiltonian
\be \CH_{BFV} = \tfrac{1}{2} \vec p^{\nabla} \cdot \vec p^{\nabla} + \frac{\vec \pi \cdot (\vec \pi \times \vec x)}{2r^2} \eta \,  \la{Hrot2}\ee
where
\be \vec p^{\nabla}= \vec p - \frac{\vec x}{r^2} \eta \CP \, . \ee 
In the  extension \eqref{Hrot1}, the last term in the decomposition \eqref{Sigmadecomp}  vanishes. In \eqref{Hrot2}, on the other hand, it is the first term that does so, with the $\eta$-coefficient 
 $\Sigma_I^{ab} \equiv \Sigma^{ab}$ taking the form 
\begin{equation}
	\Sigma_{ab} = \varepsilon_{abc}\frac{x^c}{r^2} \, .
\end{equation}





\section{Angular Momentum III: \\An infinite Koszul-Tate resolution}

Before ending this article, we want to address briefly what happens if one follows the usual BFV procedure in the singular context, not taking the shortcut by the resolution of the singular foliation as advocated here. We want to sketch this at the example of the angular momentum when not excluding the spatial origin $\vec{x}=0$. As mentioned  above, in this case, one needs to take both dependencies $\vec{x} \cdot \vec{G} = 0$ and $\vec{p} \cdot \vec{G} = 0$ into account separately. As a preparation for the  construction of the BFV charge and the BFV Hamiltonian, one now determines a Koszul-Tate (KT) resolution of the ideal generated by the constraints $G_a$. The graded coordinates introduced for this purpose, called the anti-ghosts in \cite{HT}, subsequentely serve as the momenta to the ghosts that one needs for the BFV charge. For the angular momentum, we have three anti-ghosts ${\pi}_a$ of degree -$1$, which provide the generators of the ideal after application of the KT differential:  
 $ \delta_{KT}  \left({\pi}_a \right) = G_a \, . $ For the two  dependencies between these constraints, we introduce two anti-ghosts of degree -$2$,  $\CP$ and $\bar \CP$, such that: 
 \be \delta_{KT} \CP = x^a \pi^a \: , \quad \delta_{KT} \bar \CP = p^a \pi^a \, .
 \ee 
 Note that $\delta_{KT}$ coincides with $\delta$ on the ghost momenta we introduced before, see \eqref{delta}, but now we have more of them. And it does not stop with the additional $\bar \CP$. The reducibility functions $x^a$ and $p^a$ are not independent on-shell: $x^a p^b - x^b p^a = \epsilon^{abc}G_c\approx 0$. This implies that one needs additional anti-ghots $\CP^a_3$ of degree -$3$, which then give
\be \delta_{KT} \CP^a_3 = x^a \bar \CP - p^a \CP + \tfrac{1}{2}\varepsilon_{abc} \pi^b \pi^c \, . \ee 
Here the last term is needed to ensure  $(\delta_{KT})^2=0$. However, on the space generated by $(x^a,p_a,\pi_a,\CP,\bar \CP, \CP^a_3)$  one now finds non-trivial $\delta_{KT}$-cohomology classes at degree -$3$. This in turn requires the introduction of six anti-ghosts of degree -$4$ such that 
\begin{equation}
	\delta \CP^a_4 = \varepsilon_{abc}x^b \CP_3^c + \pi^a \CP, \qquad 
	\delta \bar \CP^a_4 = \varepsilon_{abc}p^b \CP_3^c + \pi^a \bar \CP \, .
\end{equation}
 And this procedure does not stop. \footnote{We will come back to a complete description of this resolution elsewhere.} By construction, there is no cohomology of $\delta_{KT}$ except at degree zero. Therefore, with this starting point, $\CQ_{BFV}$ and the extension $\CH_{BFV}$ of \eqref{H} always exist in principle. But there is a price to be paid: the  underlying space of ghosts and anti-ghosts consists of an infinite tower of them. 
  
\section{Summary and Outlook \label{outlook}}
In this article we considered mechanical models of constrained systems 
where, as is often the case in physics, at least if the topology of spacetime  is non-trivial, the constraints are not necessarily everywhere regular and where they are permitted to have reducibilities of the first level. We provided a shortcut for the construction of a BFV-type charge $\CQ_{BFV}$, bypassing a potentially intricate Koszul-Tate procedure.  We then addressed the conditions under which an extension $\CH_{BFV}$ of the Hamiltonian exists. 
In particular, we identified the complex  \eqref{CG} governing this extension problem, providing sufficient conditions for the exitence of $\CH_{BFV}$ in this way. 

There are several challenges for future work. One  is to extend the analysis to higher level reducibilities, bringing Lie $\infty$-algebroids into play. 
Another is to derive the honest BFV extension using the Koszul-Tate algorithm for irregular constraints, even if at the cost of possibly introducing 
an infinite tower of ghosts.  Finally, one may want to see how all this plays out at the quantum level. We intend to come back to some of these points elsewhere.

\section{Acknowledgements}
We thank Camille Laurent-Gengoux and Hadi Nahari for valuable discussions and Noriaki Ikeda for making us aware of \cite{Ikedaneu}. T.S. furthermore wants to thank all of them for collaborations on related subjects, which were essential for the success of the current project. We also thank Jim Stasheff and Marc Henneaux for remarks on the manuscript. A.H.\ is grateful to the ESI in Vienna 
for support to join the program ``Higher structures and Field Theory'' as a Junior Fellow.

\section{Appendix: Angular Momentum IV}
In this appendix we want to return once more to the  example of the angular momentum in $\R^3$. While the coordinates and momenta enter the constraint surface $\vec{x} \times \vec{p} \approx 0$ symmetrically, this is not the case  for the Hamiltonian, see \eqref{3}. Now, in a region where $\vec{p} \neq 0$, excluding the origin in momentum space, one may use the BFV charge 
 \be \bar \CQ_{BFV}= \vec{\xi} \cdot (\vec{x} \times \vec{p}) + \tfrac{1}{2} 
 (\vec{\xi} \times \vec{\xi}) \cdot\vec{\pi} - \bar \eta \, \vec{p} \cdot \vec{\pi} \, ,
\ee
where we introduced a ghost-for-ghost pair $(\bar \eta,  \bar \CP)$ to implement the dependency $\vec{p} \cdot (\vec{x} \times \vec{p}) \equiv 0$. Like when excluding the spatial origin $\vec{x} \neq 0$, there is no obstruction for the BFV extension of  the Hamiltonian, but in this case it even agrees with the classical one, $\bar \CH_{BFV} = \tfrac{1}{2}\vec p \cdot \vec p$. 

One may wonder, if and how one might obtain the much more involved BFV extensions \eqref{Hrot1} and \eqref{Hrot2} from  this simple solution to the extension problem in regions where both $\vec{x}$ and $\vec{p}$ are non-vanishing. 

We only have  a partial, semi-rigorous answer to this question: Introduce a BFV charge $\CQ'$ that incorporates the two dependencies given by $\vec{x}$ and  $\vec{p}$ in a symmetrical fashion,
\be
\CQ' = \vec{\xi} \cdot (\vec{x} \times \vec{p}) + \tfrac{1}{2} 
 (\vec{\xi} \times \vec{\xi}) \cdot\vec{\pi} - \eta \, \vec{x} \cdot \vec{\pi} -\bar \eta \, \vec{p} \cdot \vec{\pi}, \label{Q'}
\ee
with now two conjugate ghost-for-ghost pairs $( \eta,  \CP)$ and $(\bar \eta,  \bar \CP)$. The charge \eqref{Q'} squares to zero in the obvious canonical bracket, $(\CQ', \CQ')' = 0$. It also agrees with a truncation of the honestly constructed BFV charge in an infinite tower expansion when following the Koszul-Tate procedure mentioned at the end of the main text above. But being such a truncation, where one drops all higher ghost-for-ghost contributions, it does not have the correct cohomology, however---therefore, the argument is only semi-rigorous at this stage.

There now is a $\CQ'$-closed extension of the classic Hamiltonian on this extended phase space and it is even globally defined: The simple
\be
\CH' = \tfrac{1}{2}\vec p \cdot \vec p - \eta \bar \CP
\la{H'}
\ee
is readily seen to satisy $(\CQ', \CH')' = 0$.

We may obtain the two sought-for BFV formulations from the above one by two different reductions in the extended phase space, where one has the canonical coordinates $(\vec{x},\vec{p},\vec{\xi},\vec{\pi},\eta,  \CP,\bar \eta,  \bar \CP)$. The first one follows from the evident choice
\be \eta := 0 \, , \;  \CP :=0. \label{red1}
\ee
It reproduces the BFV formulation $(\bar \CQ_{BFV},\bar \CH_{BFV})$ valid on regions with $\vec{p} \neq 0$.

Note that the elimination of a canonical pair, here $( \eta,   \CP)$, does not affect the brackets between the remaining variables. But one still needs to ensure that the new quantities satisfy the two master equations after constraining to~\eqref{red1}. One way of doing this is to see if they remain to have vanishing brackets when replacing the original BFV bracket $(\cdot , \cdot)'$ by a Dirac bracket (more precisely, by an adaptation of the Dirac bracket, introduced originally for second class constraints \cite{Dirac}, to the extended phase space): Restraining a graded symplectic space by putting two even functions $\alpha$ and $\beta$  which satisfy $(\alpha,\beta)'=1$ to zero, one replaces the original  bracket $(f , g)'$ between two functions $f$ and $g$ by
\be (f , g)_{D} : = (f , g)'  - (f , \alpha)' \, (\beta, g)' + (f , \beta)'\, (\alpha, g)' \, . \label{Dirac}
\ee 
Restriction to the subspace given by $\alpha=\beta =0$ and inverting the restricted graded symplectic form, in the end is reproduced by this bracket directly. Since both, \eqref{Q'} and \eqref{H'}  commute with $\eta$ in the $(\cdot , \cdot)'$ bracket,  in view of \eqref{Dirac}, it is clear that the two master equations still hold for the induced bracket when implementing \eqref{red1}.

The situation changes, however, if one wants to proceed in the same way by replacing \eqref{red1} with  the corresponding barred equations. The reason is that now $(\bar \eta,\CH')'$ does not vanish, nor does $( \CQ', \bar \CP)'$, this then would impede the vanishing of $(\CQ',\CH')_D$. To take care of this, one may search for deforming the condition $\bar \CP =0$ by $\bar \CP =F$ for some function $F$ of the unbarred variables. This procedure permits to find the function $F$ by an expansion in the ghosts, which can be chosen such that the deformed conditions become
\be \bar \eta := 0 \, , \;  \bar \CP := \frac{1}{2r^2}\left(  2\vec p \cdot \vec x \, \CP - \eta \CP^2 -\vec \pi \cdot (\vec \pi \times \vec x)\right).
\ee
With this "gauge" the two master equations remain valid and one obtains precisely the BFV data $\CQ_{BFV}$ and $\CH_{BFV})$ as given by \eqref{QBFVrotations} and \eqref{Hrot2}, respectively. The data corresponding to 
\eqref{QBFVrotations} and \eqref{Hrot1}, on the other hand, result from these by an additional canonical transformation.
\bibliographystyle{apsrev4-2.bst}
\bibliography{biblio}

\end{document}